\input{aipcheck}

\documentclass[
    ,final            
  ]
  {aipproc}

\layoutstyle{6x9}

\begin{document}

\title{Quantum Reflection of S-wave Unstable States}


\author{N. G. Kelkar}{
  address={Departamento de Fisica, Universidad de los Andes, 
Cra. 1E, 18A-10, Bogot\'a, Colombia}}

\begin{abstract}
The phase time in quantum tunneling can be disentangled into a 
dwell time plus a term arising due to the interference of the reflected 
and incident waves in front of the barrier. The interference term dominates
at low energies and as $E \rightarrow 0$, this term and hence the phase 
time becomes singular. With the $s$-wave motion in three dimensions being 
equivalent to that of a one-dimensional motion in the radial coordinate, 
a similar singularity shows up in the phase time delay of $s$-wave resonances.
Relating the scattering matrix in three dimensional collisions to the 
reflection amplitude, the interference term in tunneling can be identified
as a term given in terms of the transition matrix in scattering. Subtraction
of this term from the phase time delay gives the dwell time delay which
is finite at all energies and is useful in characterizing $s$-wave resonances
such as the $\sigma$ meson and mesic nuclei near threshold. 
\end{abstract}

\maketitle


\section{Tunneling times}
Though the tunneling of a particle through a classically forbidden region 
is one of the oldest quantum phenomenon observed, the amount of time 
spent by a particle in the barrier region has remained a controversial
topic over decades. The classical definition for the duration of 
a collision event for example is straightforward, however, 
the quantum mechanical one depends on the approach used. Hence, in an attempt
to answer the question of how long does a particle need to traverse
the barrier, several different tunneling times were defined
\cite{haugerev}. One of the earliest definition was that of a {\it phase time}  
which involves following the peak of a narrow wave packet. 
An average {\it dwell time} in scattering 
collisions was introduced by Smith \cite{smith} in 1960 and was discussed 
later in the context of one dimensional (1D) tunneling problems by B\"uttiker 
\cite{butt83}. The average was taken over reflection and transmission in 
1D and over the scattering channels in three dimensions. In 1966 
Baz' proposed a Gedanken experiment where a small magnetic field is 
confined to a region and the Larmor precession of electrons is used as a 
quantum clock. Rybachenko 
applied the method to discuss the {\it Larmor time} \cite{ryba} in 1D. 
If the electrons have a direction of polarization perpendicular 
to the direction of the field, 
the time spent in the field region was found to be 
proportional to the expectation value of a spin component. B\"uttiker 
and Landauer defined a {\it traversal time} \cite{butlan} which turns out to
be formally similar to the inverse of the 
assault frequency in a tunneling problem \cite{mehect}. 
Recently the tunneling time concepts have gained importance in the discussions 
of superluminal propagation of tunneling particles which by itself is 
a controversial topic \cite{superlum}. The Hartmann effect connected with
superluminality involves the saturation of dwell time with the barrier width. 
Since the time becomes independent of the barrier width, a thick enough 
barrier can lead to superluminal propagation.
A review and some recent applications 
of tunneling times can be found in \cite{olkhothers}.

\subsection{Dwell and phase time connection}
The dwell and asymptotic phase times are the most commonly applied time 
concepts which seem to provide reliable and complementary information 
on time aspects of tunneling processes. In a 1D treatment of 
tunneling through a barrier, following the peak of a sharp 
transmitted wave packet, $T(k)e^{i\phi_T(k)} e^{ikx-iE(k)t/\hbar}$, one 
finds \cite{haugerev} that the time difference between the arrival and 
departure of the wave packet at the barrier is given by the energy derivative
of the transmission phase. A similar analysis for the reflected wave leads 
to a reflection phase time given by the energy derivative of the reflection 
phase. A weighted sum of the two possibilities (sometimes known as group 
delay \cite{winfulprl}) is then given by $\tau_{\phi} = |T|^2 \tau_{\phi_T} 
+ |R|^2 \tau_{\phi_R}$, where $|T|^2$ and $|R|^2$ are the transmission and 
reflection probabilities, $\phi_T$ and $\phi_R$ are the transmission and 
reflection phases, 
$\tau_{\phi_T}= d\phi_T/dE$ and 
$\tau_{\phi_R}= d\phi_R/dE$. The average 
dwell time in 1D which is defined as the number of particles 
divided by the incident flux can be shown to be related to 
the above phase times. 
Such a relation for a particle with an incident energy,  
$E=\hbar^2 k^2 /2 \mu$,  
is easily obtained after some rearrangement of the 
Schr\"odinger equation and is given by \cite{haugerev,winfulprl},
\begin{equation}\label{one}
\tau_{\phi}(E) \,=\, \tau_D (E) \,-\, \hbar \,[Im(R)/ k] \,\, dk / dE\,.
\end{equation} 
The average dwell time for the barrier region extending from $x_1$ to $x_2$ 
for example is given by $\tau_D (E) = \int_{x_1}^{x_2}\, |\Psi|^2 / j$, where
$j = \hbar k/\mu$. 
The second term on the right of (\ref{one}) is the self-interference term which
arises due to the overlap of the incident and reflected waves in front of
the barrier. This term is important at low energies and 
becomes singular as $E \to 0$, thus making the phase time singular too. 
A nice demonstration of the above was done 
\cite{haugerev} for the case of an opaque rectangular barrier 
($T \ll 1$), where it was shown that the phase time and dwell times are
given as,
\begin{equation}
\tau_{\phi} \simeq {2 \,\mu \over \hbar \, \kappa \, k} \, \, \,, \, \,
\tau_D \simeq {2 \, \mu \, k\over \hbar \,\kappa \, k_0^2}, \,
\, \,{\rm with} \,
\,
\kappa^2 \,=\, k_0^2\, -\, k^2
\end{equation}
so that $\tau_{\phi} \, {\buildrel k \to 0 \over \simeq}\,\infty$ and 
$\tau_D \, {\buildrel k \to 0 \over \simeq}\,0$. 

\section{Delay times}
The tunneling times considered in the previous section represent the 
times spent by the tunneling particle interacting with the barrier. 
If one subtracts the time spent by the particle in the absence 
of the barrier, one obtains a definition for time delay. Thus, 
\begin{equation}\label{dwellphase}
\tilde{\tau}_{\phi}(E) \,=\, 
\tilde{\tau}_D (E) \,-\, \hbar \,[Im(R) / k] \,\, dk /dE\, ,
\end{equation}
where, $\tilde{\tau}_{\phi}(E)\, =\, \tau_{\phi}(E)\,-\,\tau^0(E) $ and
$\tilde{\tau}_D(E) \,=\,\tau_D(E)\,-\,\tau^0(E)$ are now the phase and the
dwell time delay respectively. 
In the early fifties, Wigner defined \cite{wigner} 
a time delay in purely elastic 
scattering collisions. Following the peak of a 
scattered wave packet, this delay was 
given by the energy derivative of the scattering phase shift. 
Considering the analogy to the definitions in 1D tunneling,
we shall refer to it henceforth as a ``phase time delay" in three dimensions 
(3D). In 3D collisions, one finds a straightforward application 
of the time delay concept for resonance 
physics \cite{usfew}. For example, if the elastic scattering
of two particles $a$ and $b$ can proceed through the formation of an on-shell
intermediate state $R$, which is formed at a time $t_1$ and decays
($R \rightarrow a + b$) at time $t_2$, then the process
$a + b \rightarrow R \rightarrow a + b$ is ``delayed'' as compared to the
non-resonant scattering process, $a + b \rightarrow a + b$, by an amount
$\Delta t = t_2 - t_1$ which corresponds to the lifetime of the state $R$. 
In the presence of inelasticities, a one
to one correspondence between time delay and the lifetime of a resonance
does not exist and one rather defines a time delay matrix \cite{smith}. 
\subsection{Density of states and dwell and phase times} 
The physical significance of the tunneling and collision times can be 
further understood through their connections to the density of states (DOS). 
A relation between the dwell time and the DOS for a 3D 
system of arbitrary shape with an arbitrary number of incoming channels
was derived in \cite{iancon}. The DOS, $\rho_{\Omega}^{3D} (E)$ 
in $\Omega$, proportional to the sum of the dwell times in 
$\Omega$ for all incoming channels was shown to be,
\begin{equation}
\rho^{3D}_{\Omega} (E) \,= \, {1 \over 2\, \pi \, \hbar}\, 
\, \sum_{n=1}^N \, \, \tau_D^n (E)\,.
\end{equation}
For the relation in 1D, the number of channels 
reduces to two and for a symmetric barrier, 
$\rho_{\Omega}^{1D} (E) = (1/\pi \hbar) \tau_D (E)$ \cite{gaspa}.

The phase time delay or rather the scattering phase shift derivative in 
Wigner's time delay, namely, 
$\tilde{\tau}_{\phi}(E) = 2 \,\hbar\, d\delta/dE$,
is related through the Beth-Uhlenbeck formula to the density of 
states \cite{beth} as,
\begin{equation}\label{bethuhl}
\,\rho_l(E)\,-\,\rho_l^0(E)\,\,=\,\biggl (\, \,{2l+1 \over \pi}\biggr )\,\, 
{d\delta_l(E) \over dE}\, , 
\end{equation}
where $\rho_l(E)$ and $\rho_l^0(E)$ are the densities of states with and without
interaction respectively. The density of states with interaction can sometimes 
be less than that without interaction. This could happen for 
repulsive potentials or due to occurrence of 
large inelasticities in elastic collisions \cite{minemegbk}. In such cases,
the right hand side of the above equation would be negative and hence 
one would obtain a negative time delay. There exists however a lower bound 
on the negativity due to causality as shown by Wigner \cite{wigner}. 
The interpretation of the phase time delay as a difference in the density of 
states however becomes problematic for the particular case of $l = 0$ as
$E \to 0$. Since the scattering phase shift,
$\delta \propto k^{2l+1}$, or rather $\delta \propto E^{l+1/2}$,  
the energy derivative $d\delta /dE \propto E^{l-1/2}$. Hence, for all 
$l \ne 0$, the phase time delay smoothly vanishes as $E \to 0$, however, 
for $l=0$, $d\delta /dE \to \infty$ as $E \to 0$. This singularity is the
same as the one which we already discussed in the previous section in 
connection with the phase time in 1D tunneling at low energies.    
In the tunneling time relation, one can subtract the interference 
term in (\ref{dwellphase}) from the phase time delay to obtain the dwell time 
delay which remains finite at all energies. Can one find an interference-like 
term in collisions to subtract from the phase time delay? To answer
this question, let us first have a look at the lifetime matrix as defined 
by Smith. 
\subsection{Multichannel delay time matrix} 
Beginning with the quantum mechanical definition of the average time of
residence, namely, the dwell time in a region, Smith defined the lifetime 
as the difference between these residence times with and without interaction. 
In an elegant derivation \cite{smith}, he found a connection between the
scattering matrix and a lifetime matrix, in 1D and 3D and for elastic 
as well as inelastic collisions.  The lifetime matrix ${\bf Q}$ is related 
to ${\bf S}$ as, ${\bf Q} \, = \, i \hbar {\bf S} d{\bf S}^{\dagger}/dE \, 
= \, - i \hbar (d{\bf S}/dE) {\bf S}^{\dagger} \, =\, {\bf Q}^{\dagger}$, such 
that ${\bf Q}$ is Hermitian. Identifying 
${\bf t} = - i \hbar \partial /\partial E$ as a time operator, he found, 
${\bf Q} = -{\bf S \, t\, S}^{\dagger} \, =\,({\bf t\, S})\,{\bf S}^{\dagger}$.
The average delay experienced by a particle injected in the $i^{th}$ channel 
was given as, $\sum_j S_{ij}^* S_{ij} \Delta t_{ij} = \Re e \,(-i \hbar \, 
\sum_j S_{ij}^* dS_{ij}/dE ) \, = Q_{ii}$, such that an element of the matrix 
$\Delta {\bf t}$ was given by, 
$\Delta t_{ij} = \Re e ( -i \hbar (S_{ij})^{-1} {dS_{ij}/dE})$,  
where $S_{ij}$ is an element of the corresponding $S$-matrix. 
If we consider a $2 \times 2 \,\, S$-matrix, such that
$S_{ii} = \eta \,e^{2i\delta _i}$
where $\delta _i$ is the real scattering phase shift for
the elastic scattering in channel $i$ and $\eta$ is the inelasticity parameter
(with $0 < \eta \le 1$), then the diagonal elements of the time delay matrix
are $\Delta t_{ii} = d\delta _i / dE $ and 
{\it have the meaning of a ``phase time delay" in a given 
channel}.  
Re-writing the ${\bf S}$ matrix as, ${\bf S} = 1 + 2\,i\,{\bf T}$, 
one can also find an expression for $\Delta {\bf t}$ in terms of ${\bf T}$. 
With $T = - (\mu k/2 \pi) t$, for the time delay in $s$-wave elastic 
collisions, a single element, $\Delta t_{ii}^{l=0}$ 
(= $\tilde{\tau}_{\phi}(E)$) is given as, 
\begin{equation}\label{tdsmat}
\tilde{\tau}_{\phi}(E) 
\,=\, {2 \hbar \over A}\,\, \biggl[\,{-\mu \over 2 \pi}\,k\,{dt_R \over dE} \,-\, {\mu^2 \,k^2 \over 
2 \pi^2}\, \biggl (\, t_I\,{dt_R \over dE}\, -\, t_R\,{dt_I \over dE}\,\biggr) \, -\, 
{\mu \over 2 \pi}\,t_R \,{dk\over dE}\,\biggr]\,,
\end{equation}
with $A = 1 \,+\, (2\mu k t_I / \pi)\,+\, (\mu^2 \,k^2 
(t_R^2\, + \,t_I^2)/\pi^2) $.
For elastic scattering in the absence of inelasticities,
the factor $A = 1$. In the above equation, $\tilde{\tau}_{\phi}(E)$ can 
once again be seen to blow up as $E \to 0$. 

Before going over to the next section, we note that the  
$S_{ij}$'s in Smith's expression are in general elements of a 
multichannel $S$-matrix, for a given partial wave $l$. 
If one does not perform a partial wave expansion, 
one obtains an energy and angle dependent
time delay of the full wave packet. A detailed analysis of various 
multichannel time delay concepts can be found in \cite{marekandme}. 

\section{The singularity and quantum reflection} 
The phase time delay which becomes singular near threshold in the case 
of $s$-wave elastic scattering poses a problem for the resonances occurring 
near threshold and the interpretation as a density of states is no more 
useful. In order to resolve the problem and subtract the singularity as in 
the case of 1D tunneling, we first notice the following: 
with there being no angle dependence of the scattering amplitude 
in the case of $s$-waves, the $s$-wave 3D motion can be viewed as a 
1D motion in the radial coordinate $r$. Having translated the problem to
a 1D one, we relate a 2-channel $S$-matrix to the reflection and transmission 
amplitudes $R(E)$ and $T(E)$ respectively. 
Considering an asymptotic wave function with incidence from the left ($L$),  
\begin{eqnarray}
\Psi_k(x)\,&=&\, e^{ikx}\,+\,\,R_L (E) \,e^{-ikx}\,\,\,\,\,\,\,\,x \, 
\rightarrow \,- \infty
\\ \nonumber
   & = & T_L(E)\,e^{ikx} \,\,\,\,\,\,\,\,\,\,\,\,\,\,
x \, \rightarrow \,+ \infty
\end{eqnarray}
and the $S$-matrix is given as, 
\begin{equation}\label{4}
{S} \,=\,\left(
\begin{array}{cc}
T_L(E)& -\,R_R(E)\\ 
\\
-\,R_L(E) & T_R(E)
\end{array}
\right )
\end{equation}
where, $T_L(E) = T_R(E) = T(E)$ 
due to time reversal invariance and $R_L(E) = R_R(E) = R(E)$ for 
symmetric potentials. Substituting for this $S$-matrix in Smith's time 
delay relation, we obtain, 
$Q_{11}\,=\,|T|^2\,\Delta t_{11}\,+\, 
|R|^2\,\Delta t_{12}$ and 
$Q_{22}\,=\,|R|^2\,\Delta t_{21}\,+\, 
|T|^2\,\Delta t_{22}$. 

Having defined the time delay matrix as above, we now resort to the 
concept of quantum reflection which corresponds to the reflection of 
a particle in a classically allowed region where there is no classical 
turning point. If we associate the amplitude $R(E)$ with such a reflection,
we can indeed use the above time delay matrix for an $s$-wave resonance
in the absence of a potential barrier too. Quantum reflection dominates 
at low energies and in {\it badlands} where the semi-classical condition
\begin{equation}
\Delta (x)\,=\, {1 \over 2 \pi}\,\,\biggl |\,{d\lambda \over dx}\,\biggr |\,
=\,\hbar\,\biggl |\,{\mu \over k^3}\,{dV \over dx}\,\biggr |\,<< 1
\end{equation}
is no more valid. Here $\lambda$ is the de Broglie wave length and $k$ the 
wave number. Since the transmission coefficient becomes negligible at low 
energies, to a good approximation one can assume that $S \to - R$ 
\cite{meprl} and with, 
$S = 1 \,-\, i \mu\,k \,(t_R \,+\, it_I)/\pi$, where
$t_R$ and $t_I$ are the real and imaginary parts of the $t$-matrix
respectively and $\mu$ is the reduced mass of the system, we obtain the
`self-interference' term of the 1D tunneling problem in terms of $t$ as
\cite{meprl},
\begin{equation}
- \hbar {Im (R) \over k}\, \,{dk \over dE}\,=\, - 
\hbar \,\mu\, {t_R \over \pi} \,\,{dk \over dE} \, .
\end{equation}
One can thus identify a dwell time delay in $s$-wave 
elastic scattering in the absence of inelasticities as, 
$\tilde{\tau}_D(E) \,=\, \tilde{\tau}_{\phi}(E)\, +\, \hbar \,
\mu\, [t_R / \pi] \,\,dk / dE$. $\tilde{\tau}_D(E)$ is finite at all energies 
and as seen before \cite{iancon} also has the interpretation of the difference 
of DOS with and without interaction. Since the reflection related term 
dominates at low energies, the phase and dwell time delay become equal 
at high energies. 
\begin{figure}
  \includegraphics[height=.29\textheight]{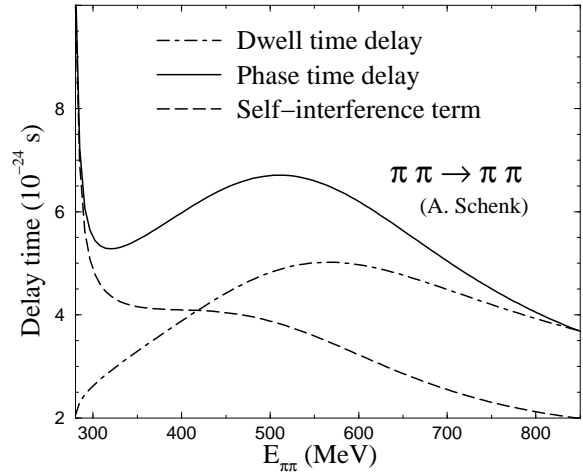}
  \caption{Delay times in $\pi \pi$ elastic scattering producing the
near threshold $\sigma$ resonance.}
\end{figure}
\subsection{Applications}
An application of this result in the search for eta-mesic 
states of nuclei was demonstrated in \cite{meprl}. 
The removal of the singularity from the phase time delay helped in 
characterizing the near threshold states of eta mesons and helium nuclei. 
Here, we show the
time delay in $\pi \pi$ elastic scattering which is evaluated using the 
parameterization of the $\pi \pi$ elastic scattering $t$-matrix as 
given in \cite{schenk}. The interpretation of the delay times in terms 
of density of states becomes important for a semi-empirical determination 
of the non-exponential decay law at large times \cite{wenonexpo}. 
The case of threshold resonances which are 
very broad is particularly interesting \cite{marektalk} and dwell time
delay can be used as an input in the evaluation of the survival probability
of the threshold resonances. As seen in Fig. 1, at large energies the 
two delay time definitions are the same.

\end{document}